\documentclass[preprint]{revtex4}

\pdfoutput=1
\usepackage{graphicx}
\usepackage{dcolumn}
\usepackage{bm}
\usepackage{amsmath}

\begin{document}

\title{Robust single-parameter quantized charge pumping}

\author{B. Kaestner}
\email[Electronic address: ]{Bernd.Kaestner@ptb.de}
\affiliation{
		Physikalisch-Technische Bundesanstalt, Bundesallee 100, 38116 Braunschweig, Germany
}

\author{V. Kashcheyevs}
\affiliation{
		Institute for Solid State Physics, University of Latvia, Riga LV-1063, Latvia
}

\author{G. Hein, K. Pierz, U. Siegner, and H. W. Schumacher}
\affiliation{
		Physikalisch-Technische Bundesanstalt, Bundesallee 100, 38116 Braunschweig, Germany.
}

\date{\today}

\begin{abstract}

This paper investigates a scheme for quantized charge pumping based on single-parameter modulation. The device was realized in an AlGaAs-GaAs gated nanowire. We find a remarkable robustness of the quantized regime against variations in the driving signal, which increases with applied rf power. This feature together with its simple configuration makes this device a potential module for a scalable source of quantized current.

\end{abstract}

\maketitle

Single electron pumps and turnstiles transporting a well defined number $n$ of charges per cycle  \cite{Averin1991} have attracted much interest, in particular for their potential application in integrated single-electron circuits \cite{nishiguchi2006} and in metrology providing a direct link between time and current units \cite{mills2006}. Different approaches have been investigated, such as arrays of gated metallic tunnel junctions \cite{geerligs1990, pothier1PBI, Keller1996, pekola2007} or semiconducting channels along which the potential can be modulated continuously \cite{Kouwenhoven1PBI, Shilton1996, fujiwara:053121, blumenthal2007a, Kaestner2007b, Fujiwara2008}. One of the main issues for applicability in metrology is to achieve a high current output simultanously with accurate charge transfer. Usually, increasing the current level by raising the frequency leads to a loss in accuracy, such that parallelization has been considered \cite{keller2000} as an alternative to faster driving. The stringent requirements on phase and amplitude matching of the driving signals typical for many systems, requiring cross-capacitance compensation for each gate-pair and channel only allow a few approaches to be considered for such a scalable current source. Here we investigate a non-adiabatic pumping scheme realized by modulating a single voltage parameter in the quantized regime \cite{Kaestner2007b, maire2008}. We find a remarkable robustness in the driving signal which should allow the application of the pump as a building block in a scalable source of quantized current.

\begin{figure}
\includegraphics{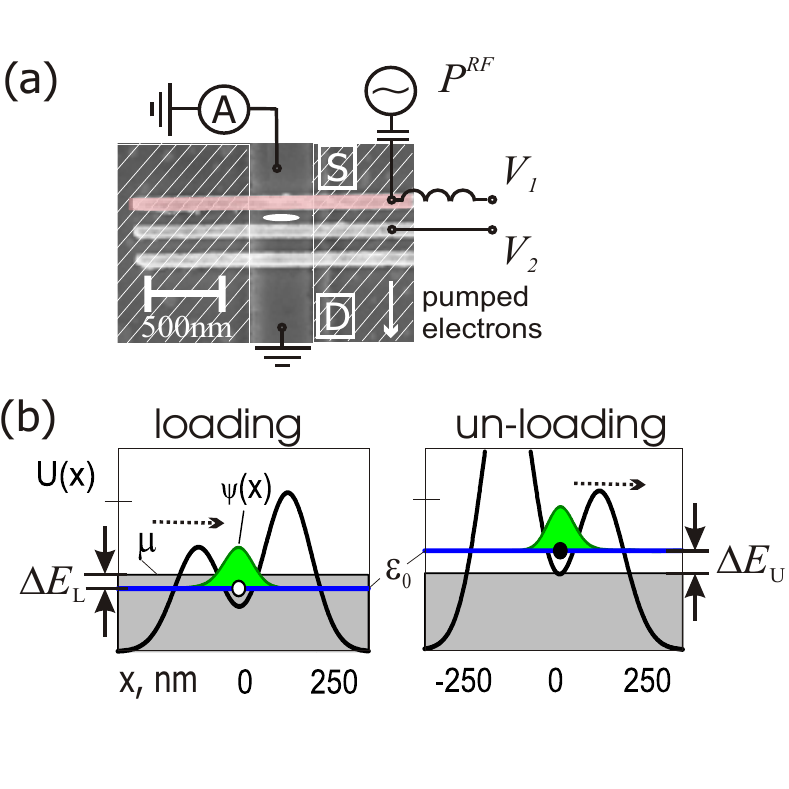}
\caption{\label{fig:device} (Color online) SEM picture of the device shown in (a). Gate voltages are indicated, showing gate 1 colored in red as being modulated. The source (S) and drain (D) reservoirs are indicated. The hatched regions are depleted of the 2D electron gas, defining a wire of about 700$\,$nm in width. A quasi-bound state is formed between gates 1 and 2, as indicated by the white ellipse. The lowest gate is not in use. (b) Schematic of the potential along the channel during loading (left) and unloading (right) of the quasi-bound state $\psi(x)$.}
\end{figure}

The device was realized in an AlGaAs/GaAs heterostructure. A 700$\,$nm wide wire connected to the two-dimensional electron gas was created by etching the doped AlGaAs layer. The device was contacted using an annealed layer of AuGeNi. This channel is crossed by three Ti-Au finger gates of 250$\,$nm separation, as shown in Fig.\ref{fig:device}(a). A quantum dot (QD) with a discrete quasi-bound state between the two upper gates is formed by applying sufficiently large negative dc voltages $V_1$ and $V_2$ to gate 1 and 2, respectively. The lowest gate was grounded and not used. An additional sinusoidal signal of power $P^{RF}$ is coupled to gate 1. If the oscillation amplitude is high enough, the energy $\epsilon_0$ of the quasi-bound state $\psi$ drops below the chemical potential $\mu$ of the leads during the first half-cycle of the periodic signal and can be loaded with an electron from the left reservoir (see Fig\ref{fig:device}(b)). During the second half-cycle $\epsilon_0$ is raised sufficiently fast above $\mu$ to avoid backtunneling and the electron can be unloaded to the right. In this way a current is driven through the sample without an applied bias and the device acts as a quantized charge pump. For details on this pumping mechanism we refer to Ref.~\cite{Kaestner2007b}.

\begin{figure}
\includegraphics{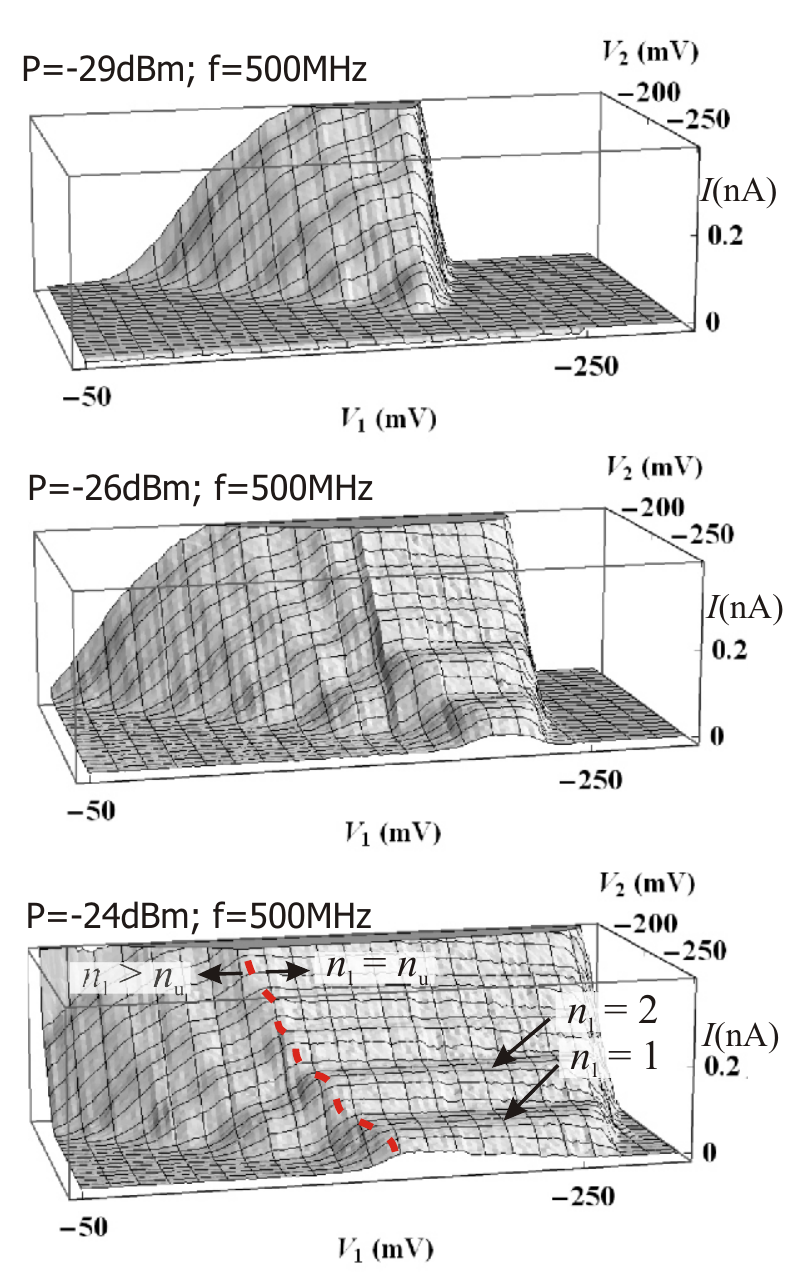}
\caption{\label{fig:mainRes} (Color online) Pumped current through the unbiased device as a function of $V_1$ and $V_2$. The rf-power of the driving signal at $f = 500\,$MHz was varied between the three cases shown.}
\end{figure}

The pumped current through the unbiased device as a function of gate voltages $V_1$ and $V_2$ is shown in Fig.~\ref{fig:mainRes}. Measurements were performed at temperature $T = 300\,$mK. Sinusoidal signals of rf-powers $P^{RF} = -29, -26, -24\,$dBm and frequency $f = 500\,$MHz were applied to gate 1. Plateaus of different quality can be seen around the values of $I = n\,e\,f = n\,80\,$pA for $n = 1, 2, 3, 4$, where on average $n$ electrons of charge $e$ per cycle are transported. To describe the behaviour qualitatively we assume that for the voltage range applied to gate 2 a high enough tunnel barrier for electrons in the drain is induced so that no electrons will be loaded from the drain. This assumption is justified since the considered voltage range lies well beyond the pinch-off voltage $V^\mathrm{po}_2 = - 100\,$mV. The step-like variation of $I$ along $V_2$ can be explained by considering that the voltage at gate 2 determines the number $n_l$ of electrons loaded \emph{from the source} in each cycle since it controls the dot potential during the loading phase (see also Fig.~\ref{fig:device}). In addition, it can prevent some of the captured electrons from being unloaded to the drain during the emission phase. The resulting current is determined by $I = n_u e f$, where $n_u \leq n_l$ is the number of unloaded electrons \emph{to the drain}. The case where $n_u < n_l$ occurs when $V_1$ is made more positive so that the rf-modulation added to $V_1$ is not sufficient to cause emission of all electrons over the barrier at gate 2. This explains the less pronounced steps along $V_1$ toward more positive values, with plateau-lengths related to the charging energy $E_C$ of the isolated dot: as soon as one electron is emitted to the drain the energy of the isolated dot is lowered by $E_C$ and might not be sufficient anymore to emit the remaining electrons over the right barrier. Tuning $V_1$ to more negative values will eventually lead to complete unloading of all loaded electrons to the drain, i.e. $n_u = n_l$. Comparing the different plateau lengths for sufficiently large power, e.g. $P^{RF} = -24\,$dBm, we conclude that for the pronounced and more extended plateau along $V_1$ one finds the case of $n_u = n_l$. The length of this plateau is a measure of the robustness of the quantized regime in the voltage applied to gate 1.

\begin{figure}
\includegraphics{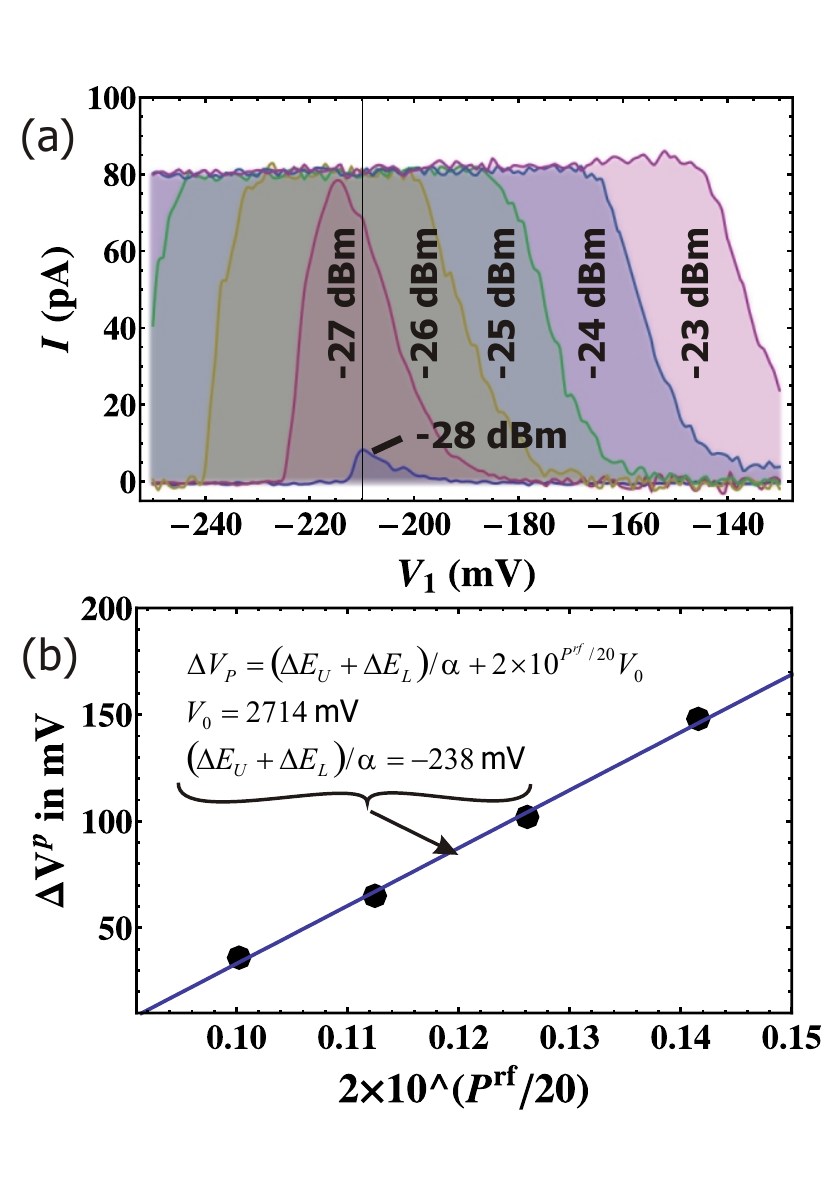}
\caption{\label{fig:Plat} (Color online) (a) Pumped current as function of $V_1$ for $V_2 = -230\,$mV different rf-powers. (b) Width of the plateau plotted for rf-powers $P = -26, -25, -24$ and $-23\,$dBm, scaled to be proportional to rf-amplitude.}
\end{figure}

To investigate the robustness further the plateau along $V_1$ at $n_u = 1$ is plotted in more detail in Fig.~\ref{fig:Plat}(a) for rf-powers $P = -28, -27, -26, -25, -24$ and $-23\,$dBm. The voltage on gate 2 was set to $V_2 = -230\,$mV. The width of the plateau increases with applied rf-power. To describe the rf-power dependence we restrict our model for simplicity to a single quasi-bound state. The energy of this state is modulated by the signal on gate 1 as $\epsilon_0(t) = E_0 + \alpha (V_1 + V^{\mathrm{rf}} \cos (2 \pi f t))$, where $\alpha < 0$ describes the conversion from voltage to energy scale, and $E_0$ is the energetic offset, including a dependence on $V_2$. Quantized pumping then requires complete loading of one electron exclusively from source and complete unloading exclusively to drain. In order for such a sequence to be possible the dot has to be isolated during the phase when $\epsilon_0$ crosses $\mu$. In terms of tunneling rates to source, $R_S$, and drain, $R_D$, we require $R_S$ and $R_D \ll f$ for $\mu - \Delta E_L < \epsilon_0 < \mu + \Delta E_U$. Here $\Delta E_U$ is the amount of energy the quasi-bound state has to gain above $\mu$ in order to unload electrons to the drain. Similarly, $\Delta E_L$ is the energy of the quasi-bound state below $\mu$ before loading sets in (see Fig.~\ref{fig:device}). This means that no electrons can be captured for $\alpha V_1 > (\mu - \Delta E_L) - (E_0 + \alpha V^{\mathrm{rf}})$. Also, no electron can be emitted for $\alpha V_1 < (\mu + \Delta E_U) - (E_0 - \alpha V^{\mathrm{rf}})$. The length of the plateau can therefore be written as $\Delta V_P = (\Delta E_U + \Delta E_L)/\alpha + 2 V^{\mathrm{rf}}$. The modulation amplitude is related to the power given in dBm via $V^{\mathrm{rf}} = 10^{P^{\mathrm{rf}}/20} V_0$, where $V_0$ corresponds to the amplitude at $P^{\mathrm{rf}} = 1\,$mW. The linear dependence of $\Delta V_P$ on $V^{\mathrm{rf}}$ is confirmed experimentally and shown in Fig. \ref{fig:Plat}(b). The line corresponds to $V_0 = 2714\,$mV and $(\Delta E_U + \Delta E_L)/\alpha = -238\,$mV. 

For future applications as a single-electron source it might also be important to determine the range of $\epsilon_0$ over which the QD is isolated. From bias-spectroscopy a value for $\alpha = -0.28\,$meV/mV has been obtained for the QD in the open regime. Assuming the same value in the isolated regime we conclude that the QD is non-adiabatically blockaded over an energy range $\Delta E_U + \Delta E_L$ of more than 50$\,$meV around $\mu$.

The accuracy of this concept demonstration device has been determined at $P=-24\,$dBm and $V_1 = -200\,$mV at the flattest part of the $n_u = 1$ plateau along $V_2$. The measured current $I = (80.0 \pm 0.5)\,$pA corresponds to the theoretical value of $e f$ to better than 1\%. In principle, the accuracy can be improved by narrowing the channel \cite{kaestner2007c}, tuning the gate width and wafer characteristics. Estimates in \cite{kaestner2007c} have shown that for suitable choice of barrier shapes an accuracy of 1 in $10^8$ could in principle be achieved.

From the investigation above we conclude that the device can be conveniently implemented into a larger network where many channels are driven by the same gate. Even if the voltage signal arriving at each channel has experienced different attenuations synchronous operation is possible in the robust high-power regime. The robustness in the driving signal and its simple configuration together with the potentially high speed of tunable barrier schemes makes non-adiabatic single-parameter pumps promising candidates for an accurately quantized, large-current source as needed for fundamental experiments in metrology and quantum electronics.

\begin{acknowledgments}
The authors acknowledge fruitful discussions with S. Amakawa and S. Lotkhov. Assistance with device fabrication from Th. Weimann, P. Hinze and H. Marx is greatly acknowledged. V.K. acknowledges financial support from the European Social Fund and the Latvian Council of Science.
\end{acknowledgments}

\end{document}